\documentclass[
  12pt,
]{article}
\usepackage{xcolor}
\usepackage[margin=1.2in]{geometry}
\usepackage{amsmath,amssymb}
\usepackage[utf8]{inputenc}
\usepackage{textcomp} 
\usepackage{lmodern}
\usepackage{caption}
\usepackage{subcaption}
\usepackage{graphicx}
\usepackage{float}
\usepackage{hyperref}
\usepackage[toc,page]{appendix}
\usepackage{authblk}

\usepackage[style=numeric]{biblatex}
\title{Applying Medical Imaging Tractography Techniques to Painterly Rendering of Images}
\author[1]{Alberto Di Biase}
\affil[1]{National Heart and Lung Institute, Faculty of Medicine, Imperial
College London, UK \protect\\
  \texttt{adibiase@ic.ac.uk}}
\date{\today}

\begin{document}
\maketitle

\begin{abstract}
Doctors and researchers routinely use diffusion tensor imaging (DTI) and
tractography to visualize the fibrous structure of tissues in the human body.
This paper explores the connection of these techniques to the painterly
rendering of images. Using a tractography algorithm the presented method can
place brush strokes that mimic the painting process of human artists,
analogously to how fibres are tracked in DTI. The analogue to the diffusion
tensor for image orientation is the structural tensor, which can provide better
local orientation information than the gradient alone. I demonstrate this
technique in portraits and general images, and discuss the parallels between
fibre tracking and brush stroke placement, and frame it in the language of
tractography. This work presents an exploratory investigation into the
cross-domain application of diffusion tensor imaging techniques to painterly
rendering of images. All the code is available at
\url{https://github.com/tito21/st-python}.
\end{abstract}

\begin{figure}[H]
  \centering
  \begin{tabular}{c c}
    \includegraphics[width=0.45\textwidth]{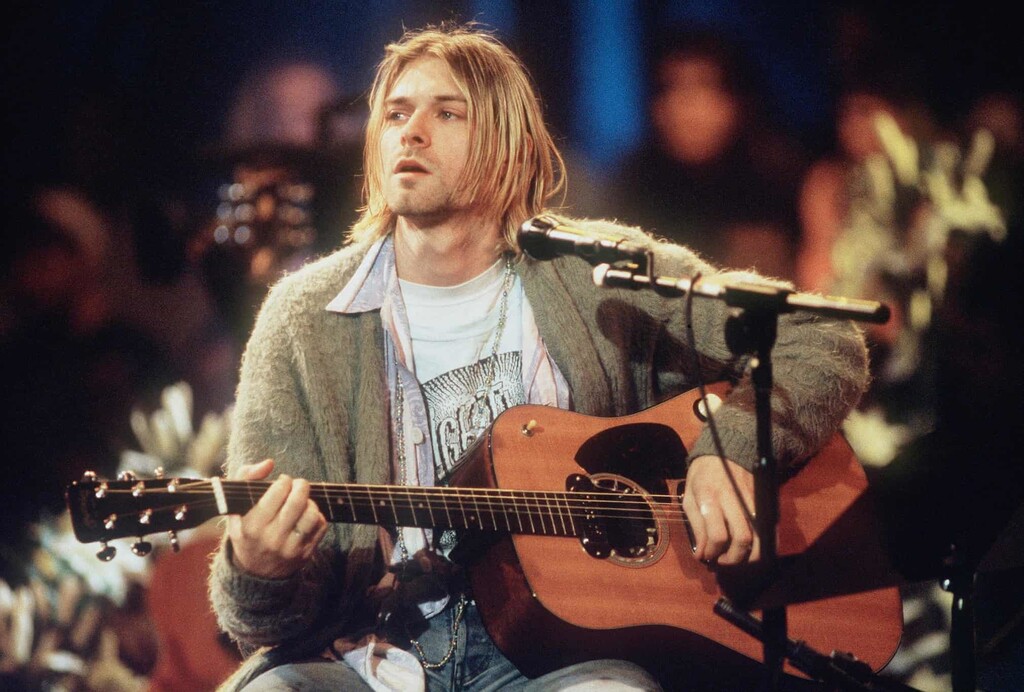} &
    \includegraphics[width=0.45\textwidth]{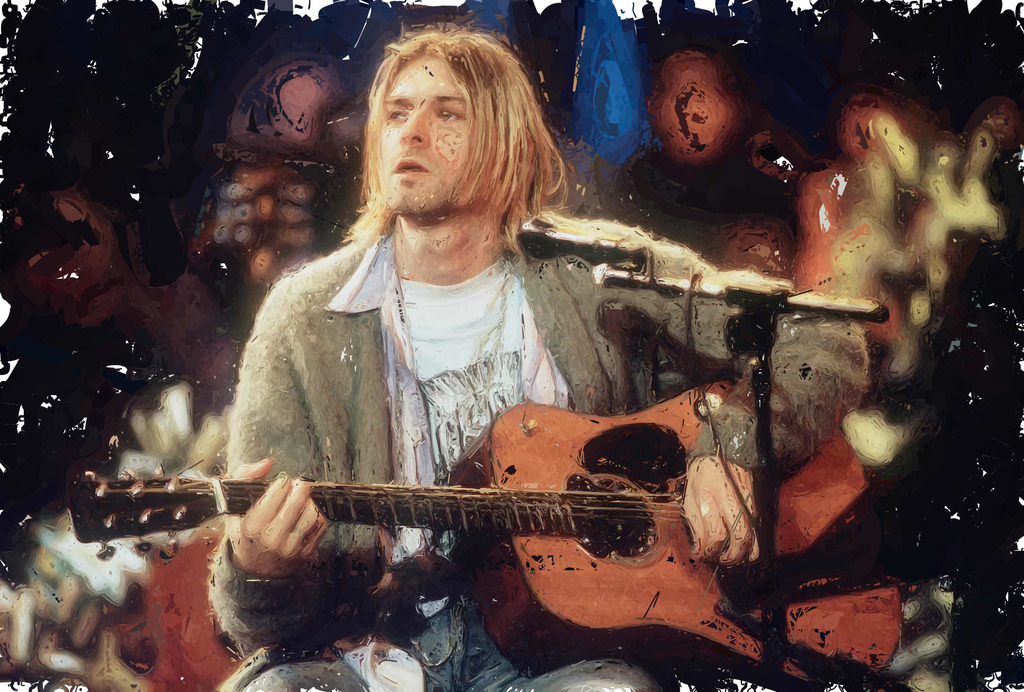}
  \end{tabular}
  \caption{Example of stylization using tractography.}
  \label{fig:stylization-example}
\end{figure}

\section{Introduction}\label{introduction}

An invaluable tool for visualizing the fibrous structure of tissue in the
human body is tractography. Tractography algorithms use the diffusion tensor to
determine the local orientation of fibres in tissue, and then place streamlines
along these orientations to create a 3D representation of the fibre pathways.
This problem is surprisingly similar to the placing coherent brush strokes in
painterly rendering of images.

This paper explores this similarity in a cross-domain framing, applying
tractography techniques to place brush strokes in images. I present the
parallels between the two problems and introduce the computer graphics community
to the mathematical framework of tractography highlighting the potential for
cross-disciplinary collaboration.

The strokes of a painter follow the contours and edges of the image subject.
This motivates the need for the extraction of gradient based quantities. In my
algorithm, this is done by the eigen decomposition of the structural tensor
(analogue to the diffusion tensor). Next, the extracted vectors are followed
with a simple tractography algorithm and then rendered with an artistic brush
stroke. An example of an image stylized with the proposed method is shown in
figure \ref{fig:stylization-example}.

The article is divided as follows: first, an overview of the diffusion tensor
imaging including a review of the tractography algorithm is given, followed by a
brief discussion of existing painterly rendering techniques. Next, a full
description of the proposed algorithm is presented. After that, results from the
NPRportrait and NPRgeneral are presented followed by discussions and
conclusions.

\subsection{Primer on Diffusion Tensor Imaging and Tractography}\label{dti-tractography}

DTI is an image modality in Magnetic Resonance Imaging (MRI) that measures the
diffusion of water across the body
\cite{odonnellIntroductionDiffusionTensor2011,basserFiberOrientationMapping}.
One of the main applications of this technique is to visually represent the
nerve tracts of the brain (and other organs), helping doctors and researchers to
study neurological disease \cite{jeurissenDiffusionMRIFiber2019}.

In DTI the acquired images are later used to fit the diffusion tensor of water
for that voxel (pixel). This is a \(3\times3\) symmetric matrix that quantifies
how much water can diffuse in each direction. Because water in biological tissue
is restricted by cell membranes the shape of the tensor is elongated in areas
were a predominant orientation is present such as in the fibre tracts. In this
case the tensor is said to be anisotropic. In the case of free water, there is
no preferred orientation and the tensor is isotropic. To measure the amount of
anisotropy in the tensor several metrics exist. The most common being the
fractional anisotropy (FA) which is a scalar value ranging from 0 (completely
isotropic) to 1 (completely anisotropic).

An eigenvalue decomposition is performed on the tensor to find the predominant
orientation. Larger eigenvalues indicate more diffusion in the direction of the
associated eigenvector. The eigensystem is generally sorted in descending order
and the first eigenvector is followed to create streamlines that represent
fibres in the tissue. This process is called tractography
\cite{odonnellIntroductionDiffusionTensor2011}.

First introduced by Basser, et al. in 2000,
\cite{basserVivoFiberTractography2000} tractography consists of following the
primary orientation of the diffusion tensor forming tracts that can be
visualized in 3D. The most basic form of this is done by solving this
differential equation:

\[
\frac{d \vec{r}(s)}{d s} = \vec{u_1}(\vec{r}(s))
\]

With \(\vec{r}(s)\) being the tract and \(\vec{u_1}\) the primary eigenvector
(in diffusion the one associated with the largest eigenvalue, and in structural
tensor the smallest). Solving this equation can be done by the well established
Runge--Kutta methods \cite{devriesFirstCourseComputational} (RK2(3) is used in
this work).

Solving the presented differential equation was the first method proposed for
tractography but more advanced techniques exist. For example, probabilistic
tractography methods that take into account uncertainty in the tensor estimation
\cite{behrensProbabilisticDiffusionTractography2007}.

\subsection{Related Works in image stylization}\label{related-work}

In the past other methods for image stylization have used the normal to the
gradient to orient the brush strokes
\cite{litwinowiczProcessingImagesVideo1997,hertzmannPainterlyRenderingCurved1998}.
These methods also achieve a painterly effect by following the direction normal
to the gradient. The algorithms to place the brush strokes are determined by a
simple procedure that can be viewed as solving the tractography differential
equation with Euler's method.

The structural tensor as a measure of orientation produces smoother brush
strokes without the need to do additional post-processing. Additionally, it is
more robust to noise and can provide more information about the image structure.
Metrics such as the coherence \cite{jahneTensorMethods1993} of the tensor can be
used to control the brush stroke width and length. In animation the structural
tensor can also be used in 3D to orient the brush strokes in a coherent manner
\cite{dahlStructureTensorComputation2019}. To the best of my knowledge the
structural tensor has not been used for image stylization in the literature.

\section{Methods}\label{methods}

\subsection{Structural Tensor}\label{structural-tensor}

For non diffusion images an analogue to the diffusion tensor is the structural
tensor
\cite{knutssonRepresentingLocalStructure2011b,bigunOptimalOrientationDetection1987}.
This tensor captures the local changes in intensity of the image. Areas with
flat texture will exhibit an isotropic tensor, while areas with edges or linear
structures will exhibit an anisotropic tensor oriented along the structure. As
with the diffusion tensor an eigenvalue decomposition is performed to extract
the local orientations.

The structural tensor is defined by the tensor that minimizes the changes in
intensities (\(V\)) when moving in an arbitrary direction. The general solution
is \(S = \sum \nabla V (\nabla V)^T\), where \(\nabla V\) is the gradient of the
image and the summation is over a neighbourhood of the image. In practice the
structure tensor is calculated using the following procedure:

\begin{enumerate}
\item
  Calculate the derivatives \(I_x\) and \(I_y\) using a Sobel filter or a
  derivative of Gaussian filter. (in this work a derivative of Gaussian with
  standard deviation \(\sigma = 1.0\) is used).
\item
  Average the products of the gradients with a Gaussian filter with a standard
  deviation \(\rho\) (in this work \(\rho = 1.0\) is used):
  (\(\overline{I_{ij}} = K_\rho * (I_i I_j)\)).
\item
  Build the tensor as:
\end{enumerate}

\[
S = \begin{bmatrix}
\overline{I_{xx}} & \overline{I_{xy}}\\
\overline{I_{xy}} & \overline{I_{yy}}
\end{bmatrix}
\]

\begin{figure}
\centering
\begin{subfigure}[t]{0.33\textwidth}
\centering
\includegraphics[width=\textwidth]{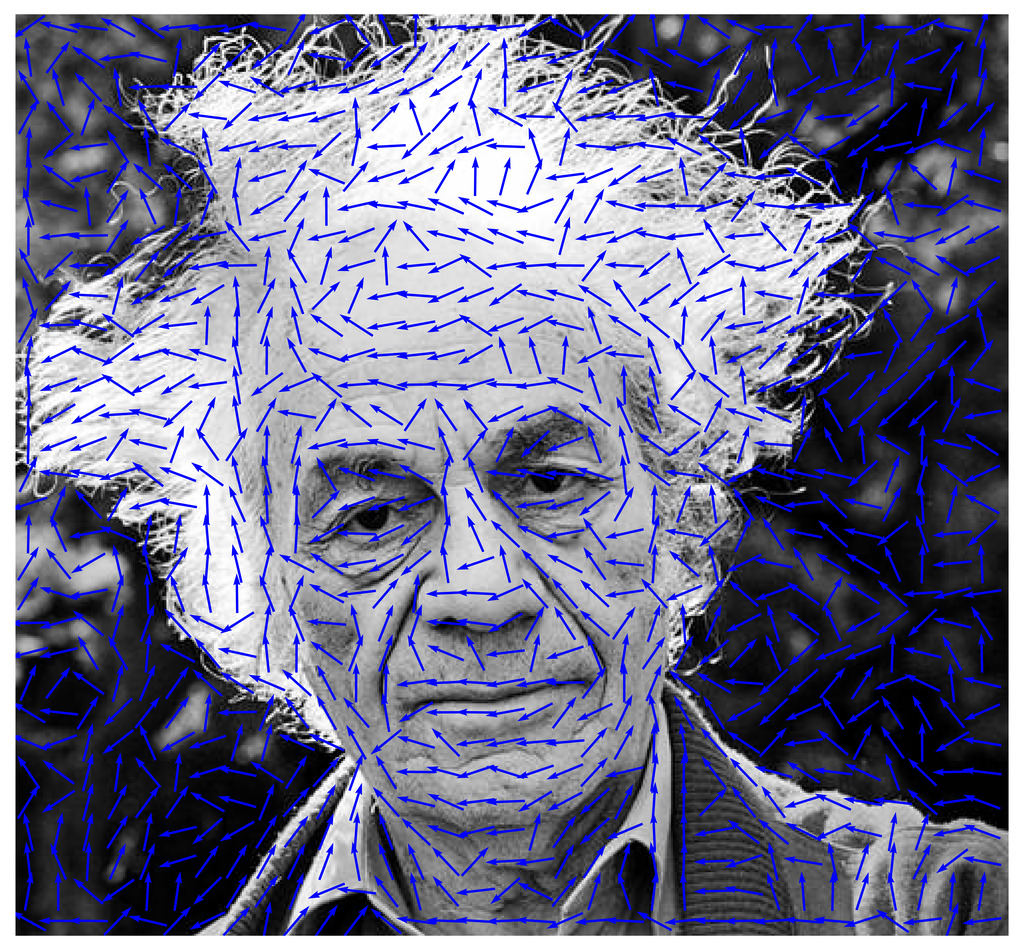}
\caption{}
\label{fig:structure-tensor-example}
\end{subfigure}
\begin{subfigure}[t]{0.33\textwidth}
\centering
\includegraphics[width=\textwidth]{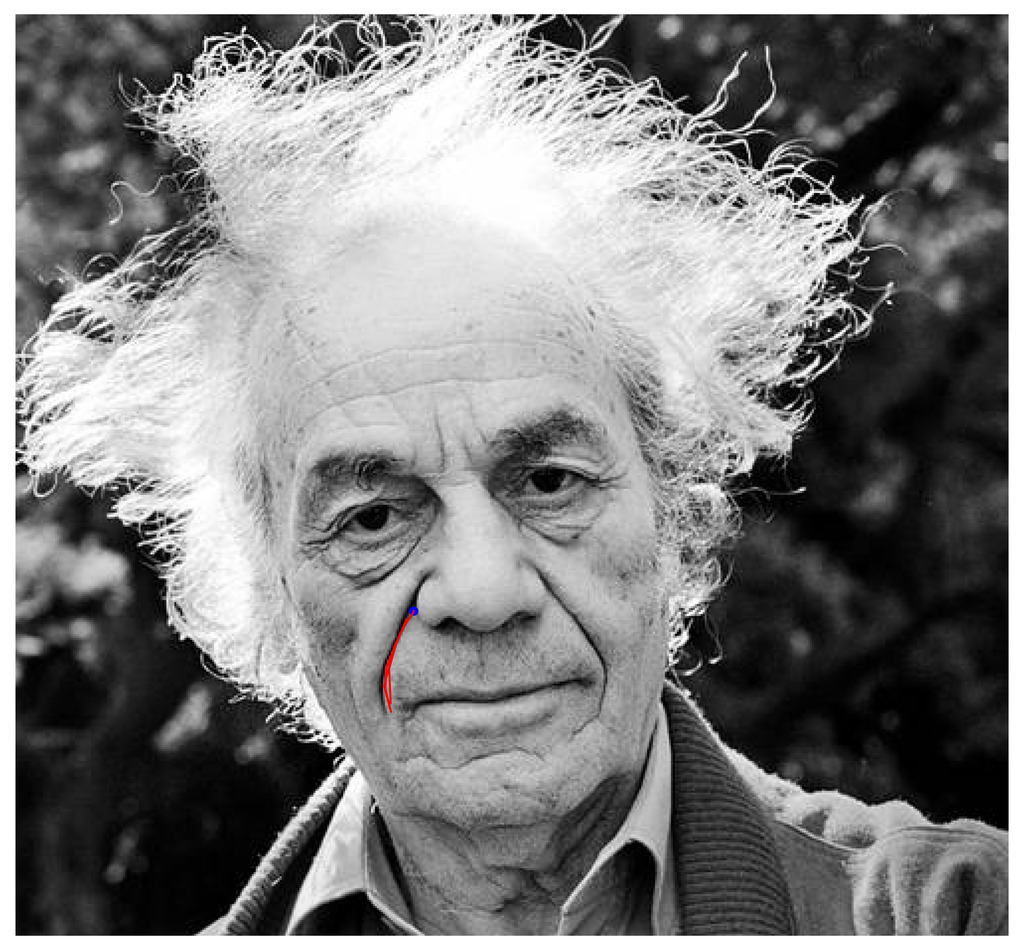}
\caption{}
\label{fig:tractography-example}
\end{subfigure}
\caption{Example of the structure tensor. The arrows are the primary eigenvector
of the structure tensor (a). Example of a tractography tract. The tract starts
from the blue point (b).}
\label{fig:structure-tensor-and-tractography}
\end{figure}

Again, if there are areas of uniform intensities in one direction, the tensor
will be oriented in that direction and is anisotropic. The other case, an
isotropic tensor would be an area with uniform intensities in all directions.
The orientation of the structures corresponds to the direction of constant
intensities (small gradients). To find the predominant orientation, the
eigenvalue decomposition is performed. However, the eigenvalues are sorted in
ascending order (reverse of the diffusion tensor) because a smaller eigenvalue
indicates the predominant orientation. This is because smaller changes in
intensity indicate that structure is present along that direction. See
\cite{dahlStructureTensorComputation2019} for more discussion on the structure
tensor. The predominant orientation is visualized in figure
\ref{fig:structure-tensor-example}. It can be seen that the arrows are aligned
with the features of the image.

\subsection{Structural Tensor vs Gradient}\label{structural-tensor-vs-gradient}

To demonstrate the improvement of the structural tensor over the gradient for
defining the orientation of the brush strokes I implemented the same filter
using the gradient. The gradient is calculated using a derivative of Gaussian
filter (with standard deviation \(\sigma = 1.0\)) and the tracts are followed in
the same way as with the structural tensor.

\subsection{Image Stylization Filter}\label{algorithm-overview}

For the generation of painterly images I use a multilayer approach
\cite{hertzmannPainterlyRenderingCurved1998} where the image is first
downsampled and lower-resolution (thicker) strokes are first rendered followed
by higher-resolution (thinner) strokes rendered on top. This follows the
painting process of most artists where the low-frequency content is painted
first, and later the details are painted on top. The following paragraphs
describe the steps for each of the layers.

First, the image converted to greyscale and is downsampled to a lower resolution
by a factor of \(\sigma\). Then the structure tensor and the associated
eigensystem is calculated. The image is divided into a grid of square cells with
the size of the brush stroke width. If the mean Euclidean distance between the
colours of the original image and the current image is above a user defined
threshold (colour threshold) in the cell a stroke is followed by the
tractography algorithm. The strokes are followed for a set length or until the
tract reaches an image border or until the coherence
\cite{jahneTensorMethods1993}, a measure of anisotropy of the local orientation,
of the pixel is below a user defined parameter (0.5 is used throughout).
Coherence for the structural tensor in 2D is defined as \(\alpha = ((\lambda_1 -
\lambda_2) / (\lambda_1 + \lambda_2))^2\) with \(\lambda_1\) and \(\lambda_2\)
the largest and smallest eigenvalues of the tensor, respectively. The order in
which the cells are drawn is set at random.

Once the tract is obtained it is simplified using the Ramer--Douglas--Peucker
\cite{ramerIterativeProcedurePolygonal1972} algorithm and converted to a Bézier
curve using the algorithm from Schneider
\cite{schneiderALGORITHMAUTOMATICALLYFITTING1990}. The colour of the stroke is
determined by the middle point of the tract. The width of the stroke is
controlled by the user. In this implementation a simple circular brush stroke
rendering is used, but any brush rendering algorithm can be used.

These steps are repeated for 4 layers with decreasing values of \(\sigma\) and
brush stroke length and width. The parameters where selected empirically for a
visually appealing result and are shown for each layer in table
\ref{tab:params}. The layers are rendered one on top of the other. Each layer
provides more detail to the image. The first layer uses large brush strokes to
render the low frequency content of the image. The number of strokes and
resolutions are increased in each layer to provide more detail.

\begin{table}
\centering
  \begin{tabular}{c c c c c}
                              & Layer 1 & Layer 2 & Layer 3 & Layer 4 \\ \hline
    Scale factor (\(\sigma\)) &      10 &       5 &       1 &     0.5 \\
    Length of strokes         &    1000 &     500 &     100 &     100 \\
    Width of strokes          &      50 &      25 &       5 &     2.5 \\
    Colour threshold          &     100 &     100 &      50 &      50 \\
  \end{tabular}
\caption{Algorithm parameters. Each layer is rendered on top of the previous
one. The scale factor is the blurring standard deviation for each level. The
length of the strokes is the maximum length of the tract to be followed in
pixels. The width of the strokes is the width of the brush stroke in pixels. The
colour threshold is the maximum error (Euclidean distance in RGB space) allowed
to render a stroke in that cell.}
\label{tab:params}
\end{table}

\subsection{Implementation Details}

A reference implementation of the algorithm is available at
\url{https://github.com/tito21/st-python} with examples of usage and the data
and code necessary to reproduce the results in this article. The code is written
in Python and uses Pycairo for rendering. To sample the vector field during the
tractography process bilinear interpolation is used. Execution time for a
\(2000\times1300\) image is approximately 15 minutes on a standard laptop
computer, though no optimization for speed was attempted.

\section{Results}\label{results}

\subsection{Structural Tensor and Gradient Comparison}\label{structural-tensor-and-gradient-comparison}

\begin{figure}
\centering
\begin{tabular}{c c}
\includegraphics[width=0.45\textwidth]{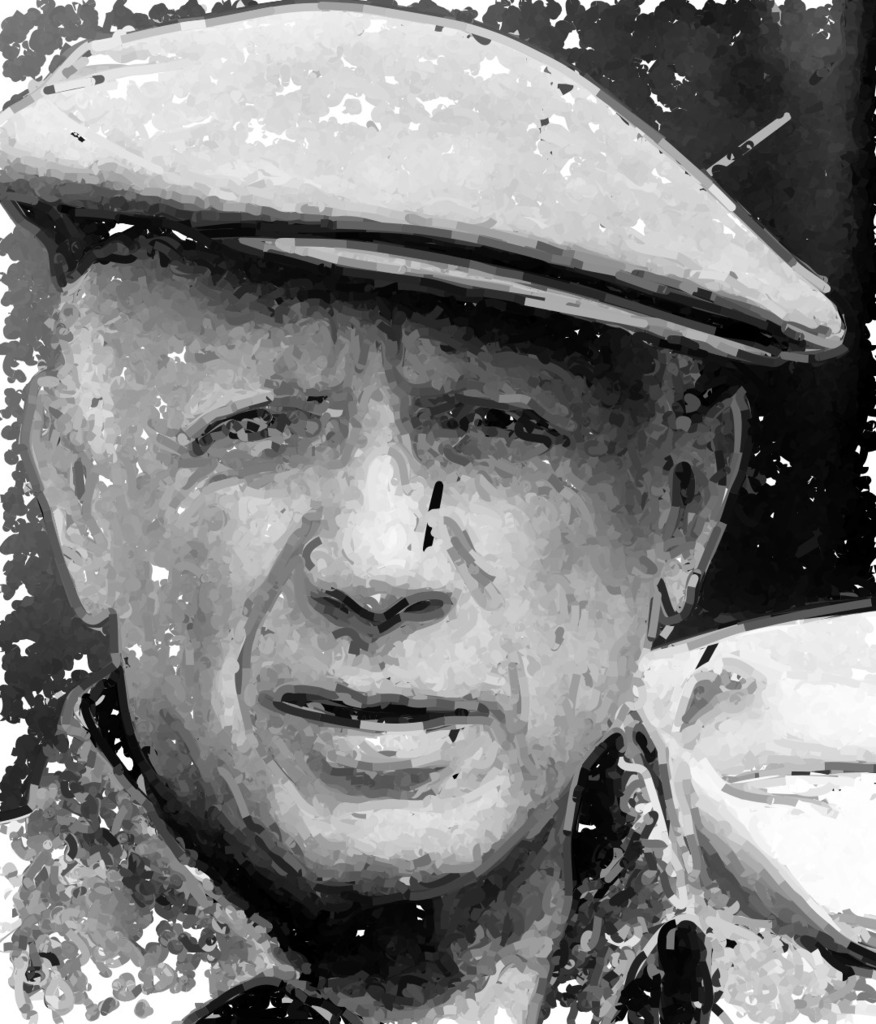} &
\includegraphics[width=0.45\textwidth]{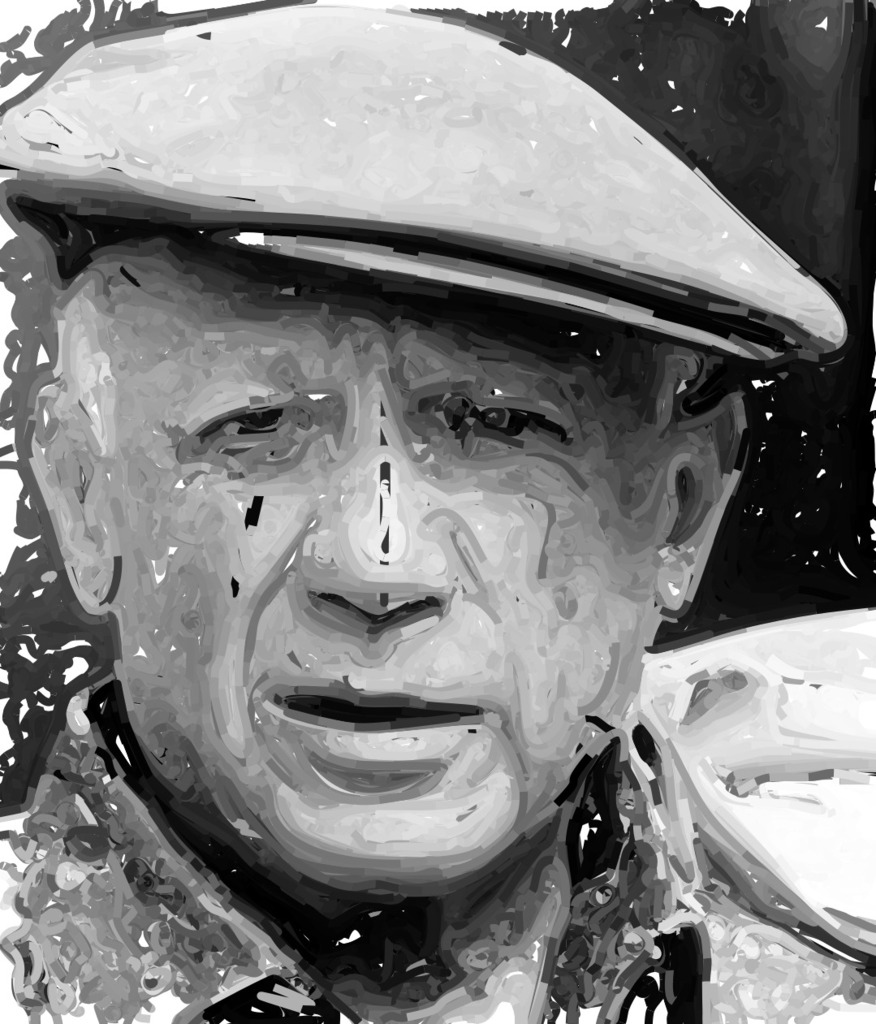} \\
\end{tabular}
\caption{Stylized images using the gradient (left) and the structural tensor
(right)}
\label{fig:gradient-vs-structural-tensor}
\end{figure}

In figure \ref{fig:gradient-vs-structural-tensor} the results of the stylization
using the gradient and the structural tensor are shown. Both images are
generated using the same parameters. The image on the left corresponds to the
gradient and the image on the right corresponds to the structural tensor. The
gradient based method produces jaggier brush strokes and results in less
clear definition of the image features. Close tracts in the structural tensor
are more aligned and produce a smoother texture.

\subsection{Evaluation Datasets}\label{evaluation-datasets}

To evaluate the proposed filter I use the NPRportrait 1.0
\cite{rosinNPRportrait10Threelevel2022} and NPRgeneral
\cite{mouldBenchmarkImageSet2016} datasets. The first dataset consists of
portraits at three levels of difficulty. The first level the subject is looking
straight into the camera with a neutral expression and with a simple background.
At the second level the images might have some background content and the
expression are not necessary neutral. In the final third level the images have
more complex lighting and backgrounds. All levels are matched by gender and
ethnicity. The NPRgeneral dataset consists of a variety of images with different
subjects and backgrounds.

For this exploratory investigation I focus on a qualitative description of the
results on the standard datasets. A formal evaluation of the performance
against other stylization methods and a more extensive ablation study of the
parameters should be conducted in future work.

\begin{figure}
\centering
\includegraphics[width=1.0\textwidth]{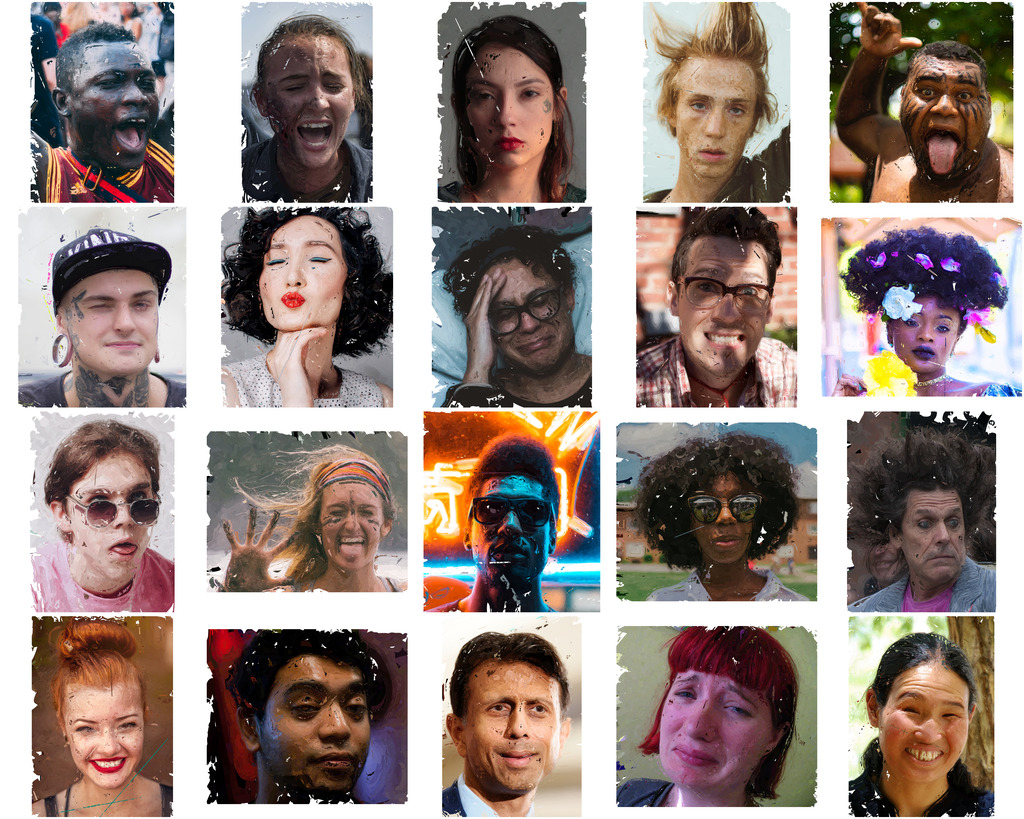}
\caption{Stylized images from NPRportrait 1.0 level 3}
\label{fig:NPRportrait3}
\end{figure}

\begin{figure}
\centering
\includegraphics[width=1.0\textwidth]{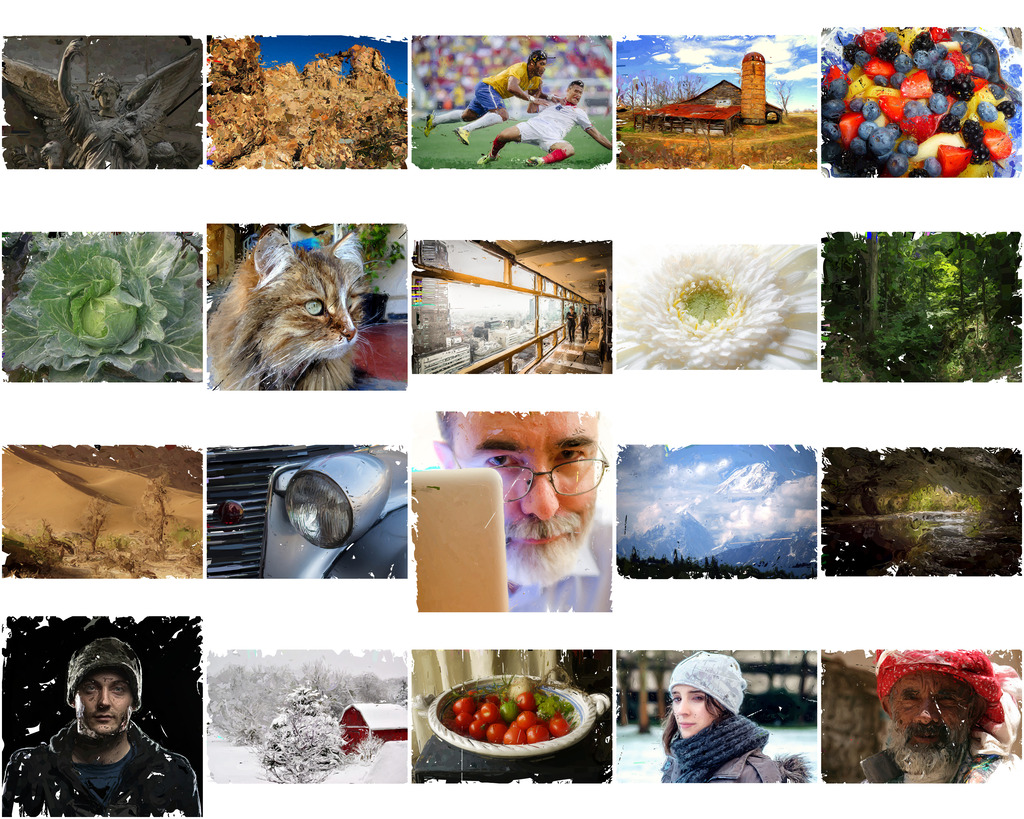}
\caption{Stylized images from NPRgeneral}
\label{fig:NPRgeneral}
\end{figure}

The stylized images from NPRportrait level 3 are shown in figure
\ref{fig:NPRportrait3}, and in figure \ref{fig:NPRgeneral} the stylized images
from NPRgeneral are shown. The reader is encouraged to view the full resolution
images in the project repository. In appendix \ref{appendix:images-org} the
original images are shown for comparison, and in appendix
\ref{appendix:images-stylized} the stylized images from the other NPRportrait
levels are shown.

Images with textures in a linear direction (for example the hair in the cat from
NPRgeneral) give the most interesting effect because each stroke follows the
hair.

In images with high frequency content or texture (for example arch in
NPRgeneral) the stylization somewhat destroys the details (however gives an
interesting effect) because they are smaller than the textures. For images with
smooth gradients (for example headlight in NPRgeneral) the stylization is not as
effective, given flat images similar to the original.

Because the colour is taken directly from the underlying image the filter
performs equally under any illumination conditions. The same applies for low
contrast situations.

Small details such as eyes are sometimes not rendered properly either because
the brush stroke is too thick or because there are no tracts in that area. The
latter problem can be solved by increasing the number of strokes in that layer
or by using a smarter algorithm to select the starting points for the tracts.

I emphasize that this work is an exploratory investigation into the cross-domain
application of diffusion tensor imaging techniques to painterly rendering and
the proposed method has not been optimized for quality or speed. Other computer
graphics experts could improve on the proposed algorithm with their expertise in
the field of non-photorealistic rendering.

\section{Discussion}\label{discussion}

\subsection{Limitations and Future Work}\label{limitations-and-future-work}

This article focused only on the selection of brush strokes from an image. The
method can select the lines to be drawn to replicate a target image by following
its contours. Once the brushes have been selected, other parameters, such as the
colour and width could be adjusted. The selection of these parameters can be
based on the target style and requires further research.

The basic implementation presented in this article preserves the original
colours of the image, but it is well known that artists play with colour for
effect. A simple randomization in the colour selection could give the desired
effect, or a more complex colour-picking selection can be used to give a
particular effect.

This implementation uses a simple brush stroke rendering that consists of
drawing simple Bézier curves between the points of the tract. More sophisticated
brush rendering algorithms can be used to give a more realistic effect
\cite{xuAdvancedDesignRealistic2003,chuRealtimePaintingExpressive2004,curtisComputergeneratedWatercolor1997}.
Additional data can be used to render the strokes such as angle and pressure can
be estimated from the tract and used to aid the brush rendering.

As in more advanced tractography algorithms, the selected strokes can be further
processed to avoid similar or overlapping strokes by removing strokes that are
too close to each other. Other post-processing of the strokes could merge
crossing strokes for an effect that would be more difficult to represent by a
real artist.

The tractography algorithm presented here corresponds with the first presented
in the literature during the early 2000s. More robust techniques have been
developed since then that improve the quality of the tracts in the presence of
noise \cite{sarwarMappingConnectomesDiffusion2019}. Also, it is common to use a
segmented approach to tractography where the image is segmented into regions and
the tracts are followed in each region. For this filter, face features can be
extracted first and then followed by tractography, ensuring that there is
appropriate detail in the most important areas.

A limitation of this work is that the strokes are selected in a greyscale
version of the image. The interaction of colours in how the tracts are selected
was not explored. Future work could include following tracts based on other
colour properties, such as hue instead of brightness, or in other colour spaces.

A final limitation is a lack of extensive algorithm evaluation. Future work
should include a formal evaluation of the performance against other stylization
methods and a more extensive ablation study of the parameters of the algorithm.

\subsection{Differences and Similarities between Medical Imaging and Painterly
Rendering}\label{differences-and-similarities}

Medical imaging tractography and stroke placement for painterly rendering are
surprisingly similar problems. In both cases the local structure of the object
under study is extracted using the eigen decomposition of a tensor (diffusion
tensor or structural tensor). Then, the primary eigenvector is followed to
create either fibre tracts or brush strokes.

However, there are some differences in the two problems. In medical imaging the
goal is to reconstruct the 3D structure of the fibres in the body as accurately
as possible, while in painterly rendering the goal is to create visually
appealing brush strokes that convey the essence of the original image. This
means that in painterly rendering there is more freedom to deviate from the
original image for artistic effect.

Interestingly the diffusion imaging community and the computer graphics
community developed similar measures of anisotropy (FA in DTI and coherence in
structural tensor analysis). The medical imaging community routinely takes
advances from computer graphics to improve their algorithms, for example when
selecting interpolation methods for vector fields, however the opposite is not
common. This work hopes to inspire researchers from both fields to collaborate
and further improve the results.

\section{Conclusion}\label{conclusion}

This work presents an exploratory investigation into applying medical imaging
concepts to the painterly rendering of images. I present the problem of brush
placement in the language of tractography, and demonstrate the use of the
structural tensor to guide the placement of brush strokes. This exploration will
hopefully inspire researchers from both fields to collaborate and further
improve the results.

\section*{Acknowledgement}

AI Chat bots were used to help with the writing of this article and in assisting
with writing the code. This work was done independently and does not represent
the views of Imperial College London.

\printbibliography

\pagebreak

\begin{appendices}

\section{Stylized images from the NPRportrait 1.0 dataset}\label{appendix:images-stylized}

\begin{figure}[H]
\centering
\includegraphics[width=1.0\textwidth]{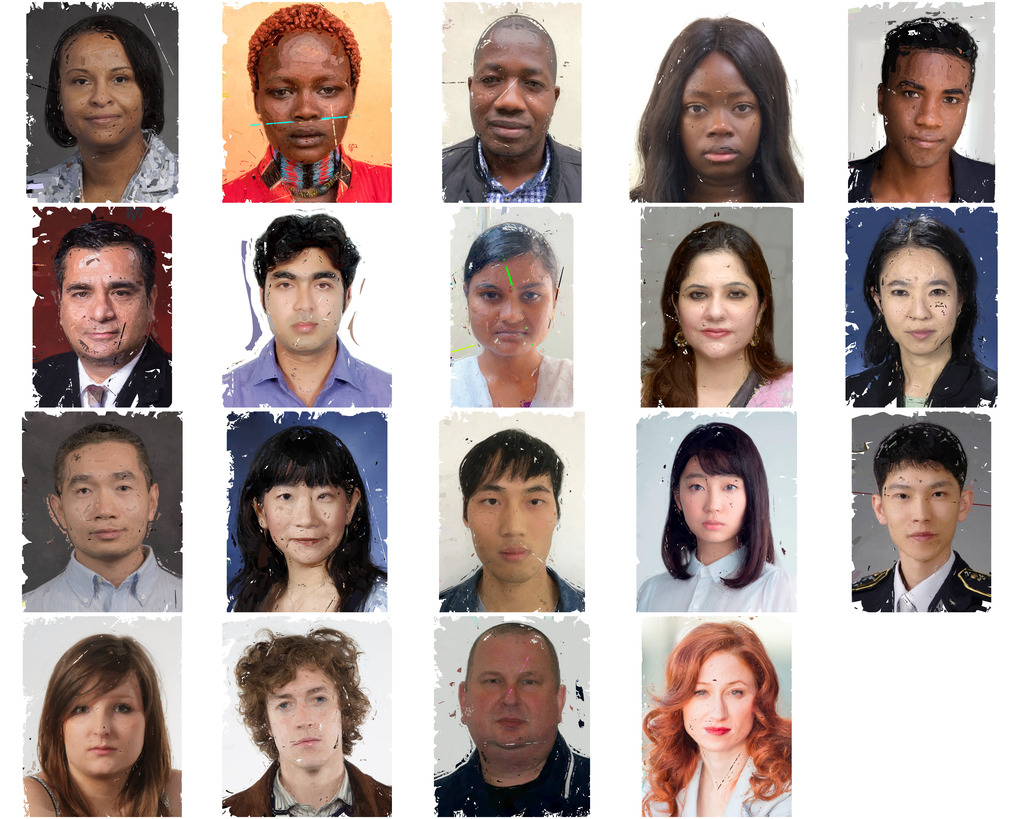}
\caption{Stylized images from NPRportrait 1.0 level 1}
\label{fig:NPRportrait1}
\end{figure}

\begin{figure}[H]
\centering
\includegraphics[width=1.0\textwidth]{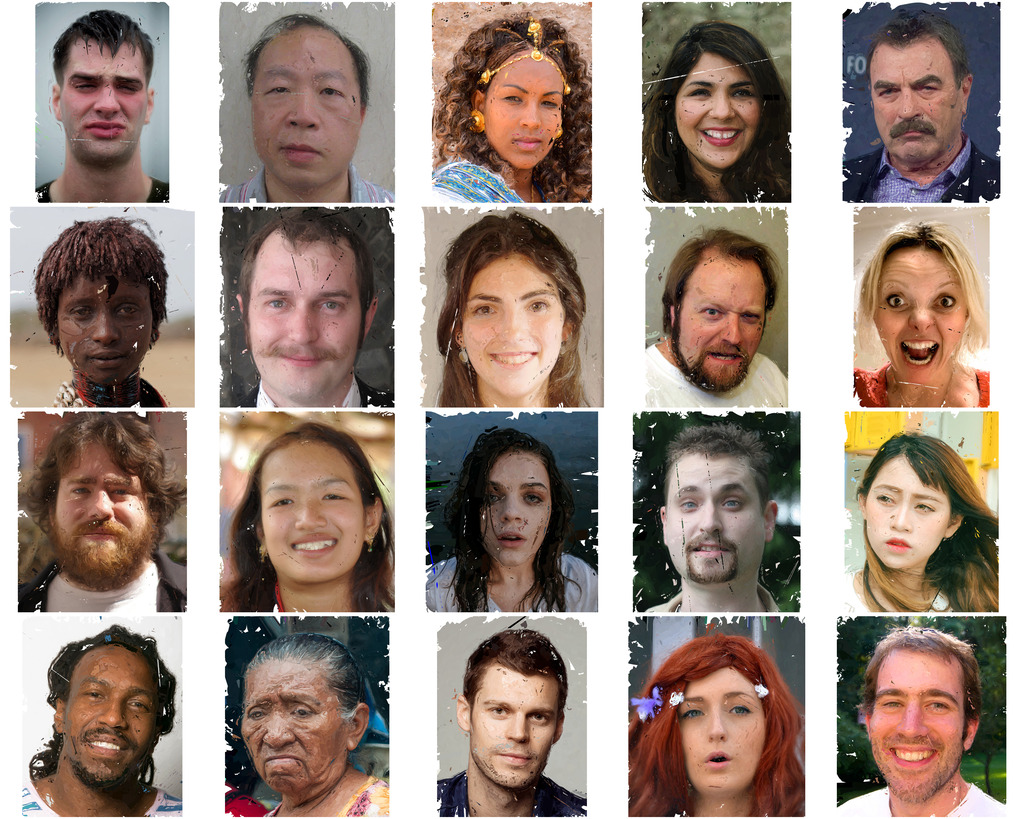}
\caption{Stylized images from NPRportrait 1.0 level 2}
\label{fig:NPRportrait2}
\end{figure}

\section{Original images}\label{appendix:images-org}

\begin{figure}[H]
\centering
\includegraphics[width=1.0\textwidth]{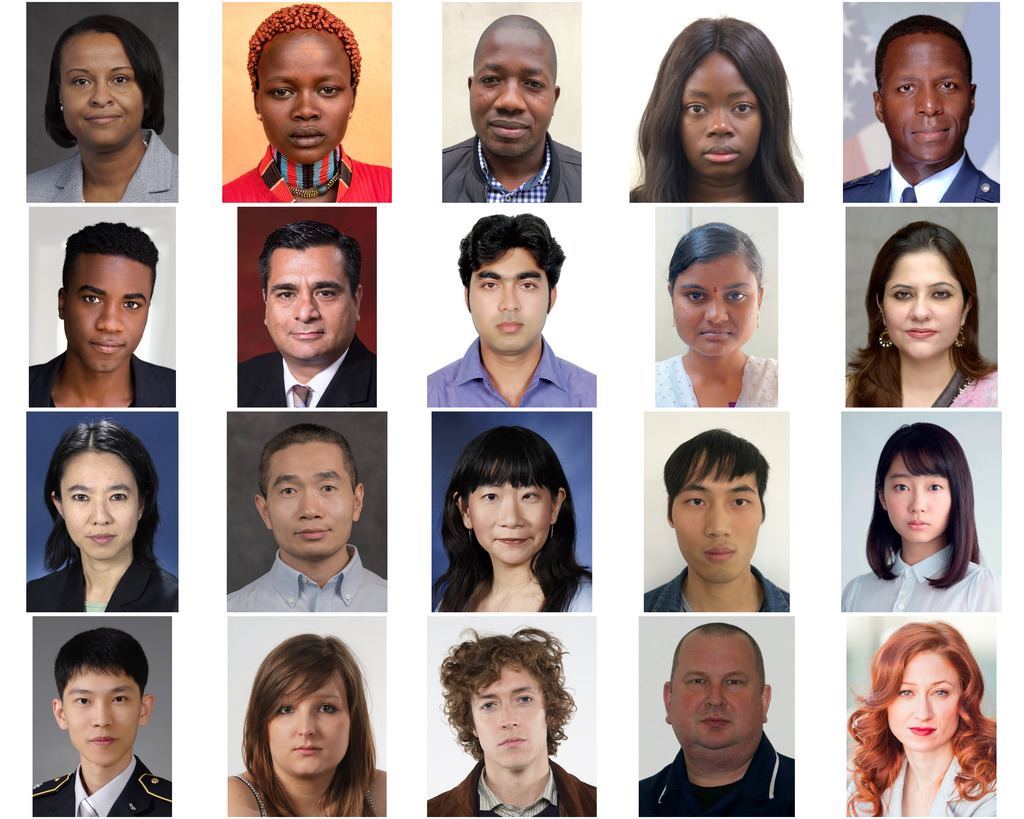}
\caption{Original images from NPRportrait 1.0 level 1}
\label{fig:NPRportrait1-orginal}
\end{figure}

\begin{figure}[H]
\centering
\includegraphics[width=1.0\textwidth]{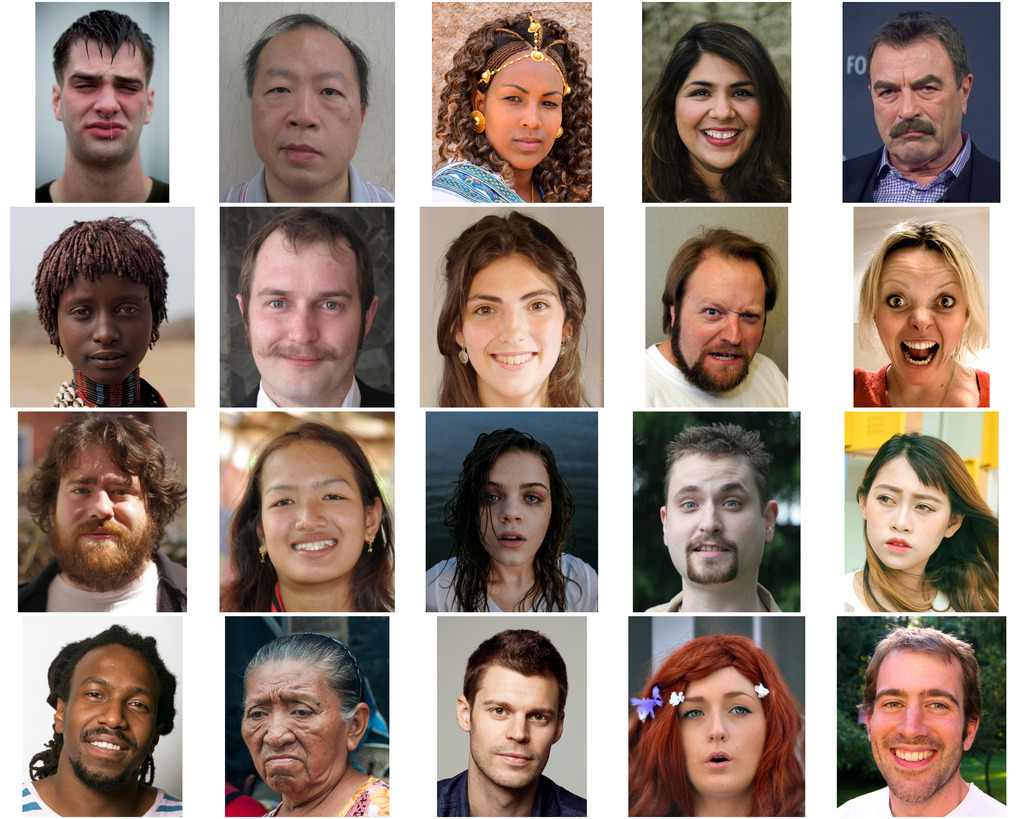}
\caption{Original images from NPRportrait 1.0 level 2}
\label{fig:NPRportrait2-orginal}
\end{figure}

\begin{figure}[H]
\centering
\includegraphics[width=1.0\textwidth]{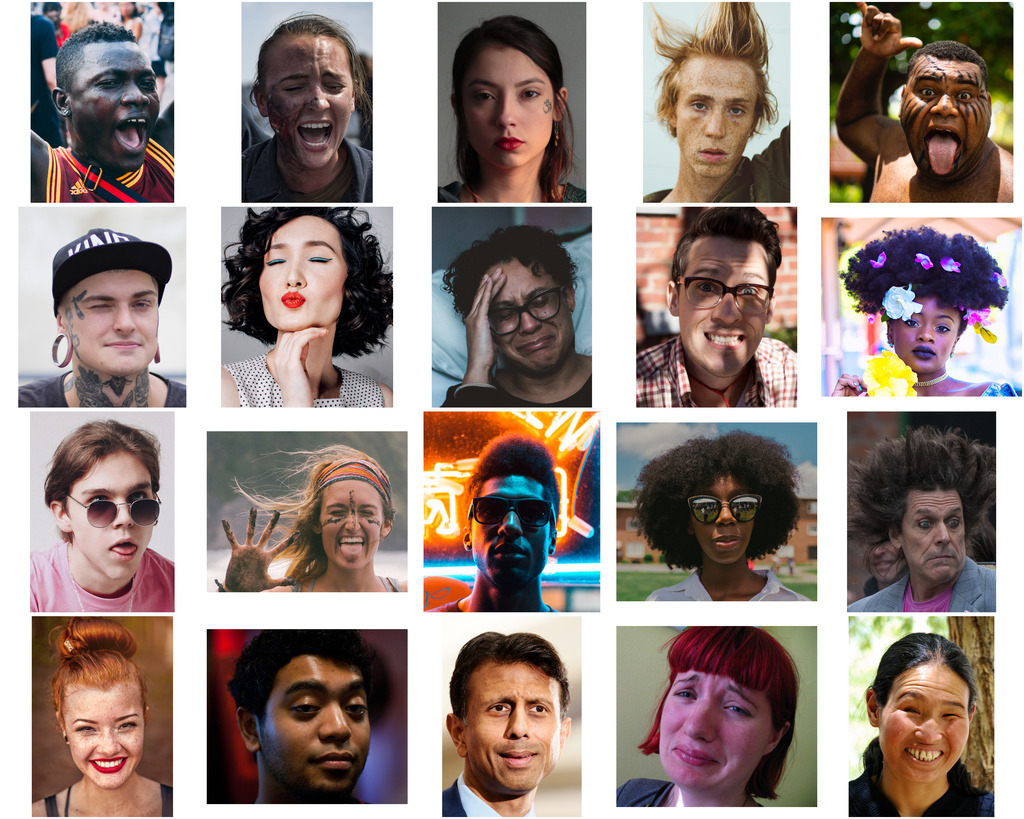}
\caption{Original images from NPRportrait 1.0 level 3}
\label{fig:NPRportrait3-orginal}
\end{figure}

\begin{figure}[H]
\centering
\includegraphics[width=1.0\textwidth]{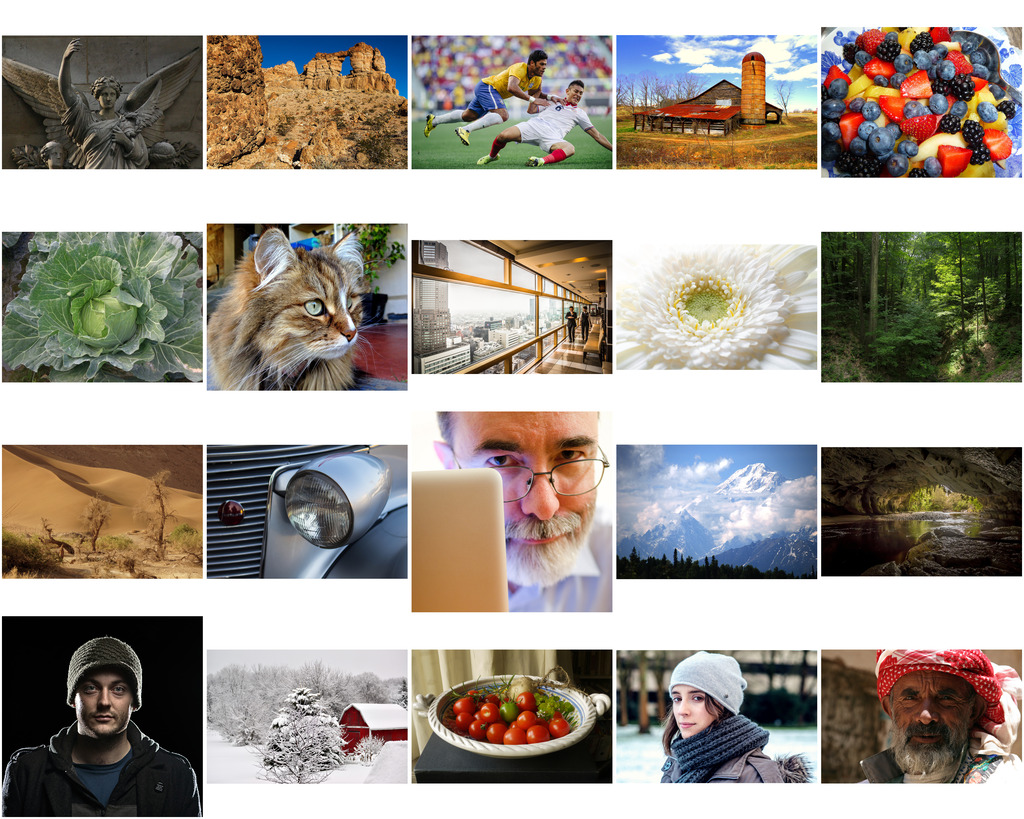}
\caption{Original images from NPRgeneral}
\label{fig:NPRgeneral-original}
\end{figure}

\end{appendices}

\end{document}